\documentclass[a4paper,12pt,review,oneside,onecolumn]{elsarticle}
\usepackage{times,amsmath,epsfig}
\usepackage{amssymb}
\usepackage{color}
\usepackage{dsfont,epstopdf}
\usepackage{graphicx}
\usepackage{bm}
\usepackage{color}
\usepackage{subfigure} 
\biboptions{numbers,sort&compress}
\usepackage{multirow} 
\usepackage{diagbox}  
\usepackage{makecell} 

\newcommand{\ba}{\begin{array}}
\newcommand{\ea}{\end{array}}
\newcommand{\be}{\begin{displaymath}}
\newcommand{\ee}{\end{displaymath}}
\newcommand{\ben}{\begin{equation}}
\newcommand{\een}{\end{equation}}
\newcommand{\bena}{\begin{eqnarray}}
\newcommand{\eena}{\end{eqnarray}}
\newcommand{\beqa}{\begin{eqnarray*}}
\newcommand{\enqa}{\end{eqnarray*}}

\newcommand{\bc}{\begin{center}}
\newcommand{\ec}{\end{center}}
\newcommand{\bi}{\begin{itemize}}
\newcommand{\ei}{\end{itemize}}
\newcommand{\benu}{\begin{enumerate}}
\newcommand{\eenu}{\end{enumerate}}
\newcommand{\bdes}{\begin{description}}
\newcommand{\edes}{\end{description}}
\newcommand{\bt}{\begin{tabular}}
\newcommand{\et}{\end{tabular}}

\newcommand \thetabf{{\mbox{\boldmath$\theta$\unboldmath}}}

\newcommand \phibf{\mbox{\boldmath$\phi$\unboldmath}}

\newcommand \jbf{{\bf j}}

\newcommand \nbf{{\bf n}}

\newcommand \sbf{{\bf s}}

\newcommand \xbf{{\bf x}}

\newcommand \zbf{{\bf z}}

\newcommand \Abf{{\bf A}}
\newcommand \Bbf{{\bf B}}

\newcommand \Hbf{{\bf H}}
\newcommand \Ibf{{\bf I}}
\newcommand \Jbf{{\bf J}}

\newcommand \Pbf{{\bf P}}

\newcommand \Rbf{{\bf R}}
\newcommand \Sbf{{\bf S}}

\newcommand \Ubf{{\bf U}}
\newcommand \Vbf{{\bf V}}
\newcommand \Wbf{{\bf W}}

\newcommand{\circlambda}{\mbox{$\Lambda$
             \kern-.85em\raise1.5ex
             \hbox{$\scriptstyle{\circ}$}}\,}

\renewcommand \thetabf{\boldsymbol{\theta}}

\renewcommand \phibf{\boldsymbol{\phi}}

\journal{Signal Processing}
\begin{document}
	\begin{frontmatter}
		\renewcommand{\thefootnote}{\fnsymbol{footnote}}
		\title{Multichannel signal detection in interference and noise when signal mismatch happens}
		\author{Weijian Liu$^{a}$,~
			Jun Liu$^{b,}$\footnote{Corresponding author. This paper was accepted by Signal Processing (Elsevier) on August 24, 2019, with DOI : 10.1016/j.sigpro.2019.107268.\\    E-mail address: liuvjian@163.com (W. Liu), junliu@ustc.edu.cn (J. Liu), ycgao@xidian.edu.cn (Y. Gao), 147616469@qq.com (G. Wang), and ylwangkjld@163.com (Y.-L. Wang).},~
			Yongchan Gao$^c$, Guoshi Wang$^a$,~Yong-Liang Wang$^a$
		}
		\address{$^a$ Wuhan Electronic Information Institute, Wuhan 430019, China}
		\address{$^b$ Department of Electronic Engineering and Information Science, University of Science and Technology of China, Hefei 230027, China} 
		\address{$^c$ Xidian University, Xi'an 710071, China}

%

\begin{abstract}
In this paper, we consider the problem of detecting a multichannel signal in interference and noise when signal mismatch happens. We first propose two selective detectors, since their strong selectivity is preferred in some situations.
However, these two detectors would not be suitable candidates if a robust detector is needed. To overcome this shortcoming, we then devise a tunable detector, which is parametrized by a non-negative scaling factor, referred to as the tunable parameter. By adjusting the tunable parameter, the proposed detector can smoothly change its capability in rejecting or robustly detecting a mismatch signal. Moreover, one selective detector and the tunable detector with an appropriate tunable parameter can provide nearly the same detection performance as existing detectors in the absence of signal mismatch. We obtain analytical expressions for the probabilities of detection (PDs) and probabilities of false alarm (PFAs) of the three proposed detectors, which are verified by Monte Carlo simulations. 
\end{abstract}


\end{frontmatter}
\section{Introduction}

Detection of a multichannel signal is a basic problem in signal processing. Many well-known detectors were proposed in the literature, such as Kelly's generalized likelihood ratio test (KGLRT) \cite{Kelly86}, adaptive matched filter (AMF) \cite{RobeyFuhrmann92}, adaptive coherent estimator (ACE) \cite{KrautScharf99}, 
and their subspace generalizations \cite{RaghavanPulsone96,PastinaLombardo01,KrautScharf01,LiuXie14b}, etc. 
The above detectors were designed without taking into account the interference, which usually exists and can significantly degrade the detection performance of a detector.
In \cite{BandieraDeMaio07a}, it is assumed that there exists interference which lies in a subspace, linearly independent of the signal subspace. This kind of interference is often referred to as subspace interference. Several detectors were proposed in \cite{BandieraDeMaio07a} in subspace interference based on the GLRT criterion. 
Recently, many other related detectors were proposed for the case of subspace interference, such as the ones in \cite{LiuLiu19TAESWald} and the references therein.

It is worth pointing out that in the above references, the signal is assumed to have an exactly known steering vector or completely lie in a given subspace. However, in practice there are many factors (e.g., not perfectly calibrated array, pointing error, and multi-path effects \cite{OrlandoRicci10,HaoLiu11}) leading to signal mismatch, for which the actual signal steering vector may not be aligned with the nominal one or not completely lie in the presumed signal subspace. 
Seldom work was done for the signal detection in the presence of interference when signal mismatch happens. A related work is \cite{LiuLiu18AES}, which analysed the statistical performance of the GLRT-based detector in \cite{BandieraDeMaio07a} in the presence of signal mismatch. However, to the best of our knowledge, no detector is specifically designed for the detection problem in interference when signal mismatch arises.

In this paper, we propose two selective (less tolerant to signal mismatch)\footnote{In some practical applications,  a selective detector would be preferred rather than a robust detector, because signal mismatch may be caused by sidelobe targets or jamming signal. More in-depth analysis can be found in \cite{PulsoneRader01}. } detectors for multichannel signal detection in the presence of interference when signal mismatch occurs. Both selective detectors have improved detection performance in rejecting mismatched signals. However, when a robust detector is needed, neither of these two detectors is a good choice. To overcome this drawback, we then design a tunable detector, which is 
parametrized by a non-negative scaling factor, called the tunable parameter. By adjusting the tunable parameter, the proposed tunable detector can flexibly control the directivity property (the capability of selectivity or robustness to mismatched signal). In particular, the tunable detector with a small tunable parameter can be much more robust to mismatched signals than existing detectors, while it, with a moderately large tunable parameter, can be more selective even than the two proposed selective detectors.
We derive analytical expressions for the probabilities of detection (PDs) and probabilities of false alarm (PFAs) of the three detectors, confirmed by Monte Carlo simulations.

The rest of the paper is organized as follows. Section 2 formulates the detection problem to be solved. Section 3 gives the proposed detectors, whose statistical properties are investigated in Section 4. Section 5 illustrates the numerical example. Finally, Section 6 concludes the paper.

\section{Problem formulation and related detectors}
For an $N\times1$ test data vector $\xbf$,\footnote{Scalars are denoted by lightfaced lowercase letters, vectors by boldfaced lowercase letters, and matrices by boldfaced uppercase letters, respectively. $\text{min}\{a,b\}$ chooses the minimum value between real numbers $a$ and $b$. $|h|$ denotes the modulus of the complex number $h$. $\text{Pr}[\cdot]$ is the probability of an event. ${\Abf^H}$ stands for the conjugate transpose of the matrix $\Abf$. $<\Abf>$ stands for the subspace spanned by the columns of $\Abf$.  
	The symbol ``$\sim$'' denotes ``be distributed as''. 
	${\cal C}{{\cal F}_{M,N}}({\kern 1pt} \xi)$ 
	and
	${\cal C}{{\cal B} _{M,N}}(\delta^2)$ denote  a complex noncentral F-distribution with $M$ and $N$ degrees of freedom (DOFs) and a complex noncentral Beta-distribution with $M$ and $N$ DOFs, respectively, and $\xi$ and $\delta^2$ are the corresponding noncentrality parameters.  When $\xi=\delta^2=0$, the two statistical distributions become central ones and written as ${\cal C}{{\cal F}_{M,N}}$ and ${\cal C}{{\cal B} _{M,N}}$, respectively. 
	Finally, ${\Ibf_N}$ is the $N \times N$ identity matrix and ${\mathbf{0}_{p \times q}}$ is the $p \times q$ null matrix.}
under signal-absent hypothesis, it consists of noise $\nbf$ and interference $\jbf$. In contrast, under signal-present hypothesis, $\xbf$ contains noise $\nbf$, interference $\jbf$, and useful signal $\sbf$. The interference $\jbf$  and signal $\sbf$ are assumed to lie in known linearly independent subspaces but with unknown coordinates. Precisely,  $\jbf$ and $\sbf$ can be expressed as  $\jbf=\Jbf\phibf$ and $\sbf=\Hbf\thetabf$, respectively. The $N\times p$ full-column-rank matrix $\Hbf$ spans the signal subspace, while the $N\times q$ full-column-rank matrix $\Jbf$ spans the interference subspace. The $q\times 1$ vector $\phibf$ and $p\times 1$ vector $\thetabf$ denote the interference and signal coordinates, respectively. Note that $p+q\le N$, due to the assumption of linear independence of the interference subspace and signal subspace. The noise $\nbf$ is Gaussian distributed, with a zero mean and a covariance matrix $\Rbf$, which is usually unknown in practice. To estimate $\Rbf$,  it is assumed that there are $L$ noise-only independent and identically distributed (IID) training data, denoted as $\xbf_l$, $l=1, 2, \cdots, L$
, sharing the same covariance matrix with the test data. Thus, the binary hypothesis test to be solved is summarized as
\begin{equation}
\label{1}
\left\{ \begin{array}{l}
{\text{H}_0}:\xbf =  \Jbf\phibf  + \nbf,
{\kern 1pt} {\kern 1pt} {\kern 1pt}
{\xbf_l} = {\nbf_l},\ l=1, 2, \cdots, L, \\
{\text{H}_1}:\xbf = \Hbf\thetabf + \Jbf\phibf + \nbf,
{\kern 1pt} {\kern 1pt} {\kern 1pt}
{\xbf_l} = {\nbf_l},\ l=1, 2, \cdots, L,
\end{array} \right.
\end{equation}
where 
$\nbf_l$ is the noise in the $l$th training data vector $\xbf_l$.

For the detection problem in \eqref{1}, the GLRT and two-step GLRT (2S--GLRT) are \cite{BandieraDeMaio07a}
\begin{equation}
\label{GLRTHE}
t_\text{GLRT--I}=\frac{\tilde\xbf^H\Pbf_{\Pbf_{\tilde\Jbf}^\bot\tilde\Hbf}\tilde\xbf}
{1+\tilde\xbf^H\Pbf_{\tilde\Jbf}^\bot\tilde\xbf-\tilde\xbf^H\Pbf_{\Pbf_{\tilde\Jbf}^\bot\tilde\Hbf}\tilde\xbf}
\end{equation}
and
\begin{equation}
\label{2SGLRTHE}
t_\text{2S--GLRT--I}= {\tilde\xbf^H\Pbf_{\Pbf_{\tilde\Jbf}^\bot\tilde\Hbf}\tilde\xbf},
\end{equation}
respectively, where
\begin{equation}
\label{xbf_tilde}
\tilde \xbf = {\Sbf^{ -1/2}}\xbf, \ \ \tilde \Jbf = {\Sbf^{ - 1/2}}\Jbf, \ \ \tilde \Hbf = {\Sbf^{ -1/2}}\Hbf,
\end{equation}
\begin{equation}
\label{Pb_Pj}
\Pbf_{\tilde \Jbf}^ \bot  = {\Ibf_N} - {\Pbf_{\tilde \Jbf}}, \ \
{\Pbf_{\tilde \Jbf}} = \tilde \Jbf{({\tilde \Jbf^H}\tilde \Jbf)^{ - 1}}{\tilde \Jbf^H},
\end{equation}
\begin{equation}
\label{Pjvh}
{\Pbf_{\Pbf_{\tilde \Jbf}^ \bot \tilde \Hbf}} = \Pbf_{\tilde \Jbf}^ \bot \tilde \Hbf{({\tilde \Hbf^H}\Pbf_{\tilde \Jbf}^ \bot \tilde \Hbf)^{ -1}}{\tilde \Hbf^H}\Pbf_{\tilde \Jbf}^ \bot,
\end{equation}
and $\Sbf =\sum_{l=1}^{L}\xbf_l\xbf_l^H$
is $L$ times the sample covariance matrix (SCM). For convenience, the detectors in \eqref{GLRTHE} and \eqref{2SGLRTHE} are referred to as the GLRT with interference rejection (GLRT--I) and 2S--GLRT with interference rejection (2S--GLRT--I), respectively.

To the best of our knowledge, no detector is specifically designed for the detection problem in \eqref{1} when signal mismatch happens.

\section{Proposed detectors}
In this section we first propose two selective detectors for mismatched signals, and then propose a tunable detector, which can smoothly adjust its detection performance for mismatched signals.


It is observed that \eqref{GLRTHE} and \eqref{2SGLRTHE} have similar forms as the subspace-based GLRT (SGLRT) \cite{RaghavanPulsone96,PastinaLombardo01} and subspace-based AMF (SAMF) \cite{LiuZhang12b}, respectively\footnote{This would be more obvious if we introduce the quantities $\tilde{\zbf}=\Pbf_{\tilde\Jbf}^\bot\tilde{\xbf}$ and $\tilde{\Abf}=\Pbf_{\tilde\Jbf}^\bot\tilde{\Hbf}$ and substituting them into \eqref{GLRTHE} and \eqref{2SGLRTHE}.}.
The SGLRT and SAMF were designed without taking the possibility of signal mismatch, and they have poor detection performance in terms of rejecting mismatched signals.
Two well-known selective detectors for mismatched signals in the absence of interference are the adaptive beamformer orthogonal rejection test (ABORT)  \cite{PulsoneRader01} and whitened ABORT (W--ABORT) \cite{BandieraBesson07TSP_WABORT}.
According to the detection statistics of the ABORT and W--ABORT, we can analogously design the following two selective detectors in the presence of interference
\begin{equation}
\label{D_ABORT}
t_\text{ABORT--I}=\frac{1+\tilde\xbf^H\Pbf_{\Pbf_{\tilde\Jbf}^\bot\tilde\Hbf}\tilde\xbf}
{1+\tilde\xbf^H\Pbf_{\tilde\Jbf}^\bot\tilde\xbf-\tilde\xbf^H\Pbf_{\Pbf_{\tilde\Jbf}^\bot\tilde\Hbf}\tilde\xbf}
\end{equation}
and
\begin{equation}
\label{D_WABORT}
t_\text{W--ABORT--I}=\frac{1+\tilde\xbf^H{\Pbf_{\tilde\Jbf}^\bot}\tilde\xbf}
{(1+\tilde\xbf^H\Pbf_{\tilde\Jbf}^\bot\tilde\xbf-\tilde\xbf^H\Pbf_{\Pbf_{\tilde\Jbf}^\bot\tilde\Hbf}\tilde\xbf)^2},
\end{equation}
which, for convenience, are referred to as the ABORT with interference rejection (ABORT--I) and W--ABORT with interference rejection (W--ABORT--I), respectively.  

It is expected that the proposed ABORT--I and W--ABORT--I can provide better performance in terms of rejecting mismatched signals. In fact, this is indeed the case, as shown in Section 4 below. However, they suffer from performance loss if a robust detector is needed. 
To cope with this problem, we introduce the following tunable detector
\begin{equation}
\label{D_TWABORT}
t_\text{T--W--ABORT--I}=\frac{1+\tilde\xbf^H{\Pbf_{\tilde\Jbf}^\bot}\tilde\xbf}
{(1+\tilde\xbf^H\Pbf_{\tilde\Jbf}^\bot\tilde\xbf-\tilde\xbf^H\Pbf_{\Pbf_{\tilde\Jbf}^\bot\tilde\Hbf}\tilde\xbf)^\kappa},
\end{equation}
which is named as the tunable W--ABORT--I (T--W--ABORT--I). The non-negative factor $\kappa$ is taken as the tunable parameter.

Roughly speaking, the numerator of \eqref{D_TWABORT} collects the total energy of the quasi-whitened test data $\tilde\xbf$ after interference suppression\footnote{Quasi-whitening is done by multiplying the test data $\xbf$ with $\Sbf^{-\frac{1}{2}}$, and  interference suppression is owing to multiplying the quasi-whitened test data $\tilde\xbf$ with the orthogonal projection matrix ${\Pbf_{\tilde\Jbf}^\bot}$.}. 
In contrast, the denominator of \eqref{D_TWABORT} gathers 
the energy of the quasi-whitened test data $\tilde\xbf$ projected onto the subspace orthogonal to the signal-plus-interference\footnote{This is more evident if we rewrite 
$\tilde\xbf^H\Pbf_{\tilde\Jbf}^\bot\tilde\xbf-\tilde\xbf^H\Pbf_{\Pbf_{\tilde\Jbf}^\bot\tilde\Hbf}\tilde\xbf$	as $\tilde\xbf^H\Pbf_{{\tilde\Bbf}}^\bot\tilde\xbf$, where $\tilde{\Bbf}=[\tilde{\Jbf},\tilde{\Hbf}]$.}.
Hence, by adjusting the tunable parameter $\kappa$, one can control the directivity property of the T--W--ABORT--I for mismatched signals. Increasing $\kappa$ will make the T--W--ABORT--I more and more selective, while decreasing $\kappa$ will make the T--W--ABORT--I more and more robust.

In particular, the T--W--ABORT--I with $\kappa=0$ is most robust to signal mismatch, and in this case the T--W--ABORT--I reduces
\begin{equation}
\label{D_WABORT0x}
t_{\text{W--ABORT--I},\kappa=0} =\tilde\xbf^H{\Pbf_{\tilde\Jbf}^\bot}\tilde\xbf,
\end{equation}
where the constant is ignored.
Equation \eqref{D_WABORT0x} as be recast as
\begin{equation}
\label{D_WABORT0xx}
t_{\text{W--ABORT--I},\kappa=0}^\prime=\tilde\zbf^H\tilde\zbf,
\end{equation}
which has the same form as the adaptive energy detector (AED) in \cite{RaghavanQiu95}. In \eqref{D_WABORT0xx}, $\tilde{\zbf}=\Pbf_{\tilde\Jbf}^\bot\tilde{\xbf}$.
When $\kappa=2$, the T--W--ABORT--I reduces to the W--ABORT--I. When $\kappa=1$, the T--W--ABORT--I reduces to
\begin{equation}
\label{D_WABORT1}
t_{\text{W--ABORT--I},\kappa=1}=\frac{1+\tilde\xbf^H{\Pbf_{\tilde\Jbf}^\bot}\tilde\xbf}
{1+\tilde\xbf^H\Pbf_{\tilde\Jbf}^\bot\tilde\xbf-\tilde\xbf^H\Pbf_{\Pbf_{\tilde\Jbf}^\bot\tilde\Hbf}\tilde\xbf},
\end{equation}
which is equivalent to the GLRT--I, since $t_{\text{W--ABORT--I},\kappa=1}=1/(1-t_\text{GLRT--I})$ can serve as a monotonically increasing function of $t_\text{GLRT--I}$.

\section{Statistical performance of the proposed detectors in the presence of signal mismatch}
When signal mismatch happens, the actual signal, denoted as $\sbf_0$, will not belong to the nominal signal subspace $<\Hbf>$.
To facilitate the derivations of the statistical properties of the proposed detector, a loss factor is introduced
\begin{equation}
\label{Beta_HE}
{\beta} = {(1 + {\tilde \xbf^H}\Pbf_{\tilde \Jbf}^ \bot \tilde \xbf - {\tilde \xbf^H}{\Pbf_{\Pbf_{\tilde \Jbf}^ \bot \tilde \Hbf}}\tilde \xbf)^{ - 1}}.
\end{equation}
Using \eqref{GLRTHE} and \eqref{Beta_HE}, we can rewrite \eqref{D_ABORT}, \eqref{D_WABORT}, and \eqref{D_TWABORT} as
\begin{equation}
\label{ABORTeq}
t_\text{ABORT--I}  = {t_\text{GLRT--I}} + {{\beta }},
\end{equation}
\begin{equation}
\label{WABORTeq}
t_\text{W--ABORT--I}  = (1+{t_\text{GLRT--I}})\beta ,
\end{equation}
and
\begin{equation}
\label{TWABORTeq}
t_\text{T--W--ABORT--I}  = \beta ^{\kappa-1}(1+{t_\text{GLRT--I}}) ,
\end{equation}
respectively.

Using \eqref{ABORTeq}
-\eqref{TWABORTeq}, 
we can readily obtain the expressions for the conditional PDs and PFAs of the three proposed detectors, conditioned on $\beta$.
Precisely, the conditional PDs of the ABORT--I, W--ABORT--I, and T--W--ABORT--I can be expressed as
\begin{equation}
\label{PDABORT}
\begin{array}{c}
\begin{aligned}
\text{PD}_{{\text{ABORT--I}}|\beta} = \text{Pr}[{t_\text{GLRT--I}} + {{\beta }} > {\eta _{\text{a}}};{\text{H}_1}]
 =1 -{{\cal P}_{1}{({\eta _{\text{a}}}-\beta)}},
\end{aligned}
\end{array}
\end{equation}
\begin{equation}
\label{PDWABORT}
\begin{array}{c}
\begin{aligned}
\text{PD}_{{\text{W--ABORT--I}}|\beta}= \text{Pr}[(1+{t_\text{GLRT--I}})\beta > {\eta _{\text{w}}};{\text{H}_1}]
= 1 -{{\cal P}_{1}{\left( \frac{\eta _{\text{w}}}{\beta}-1   \right)}},
\end{aligned}
\end{array}
\end{equation}
and
\begin{equation}
\label{PDTWABORT}
\begin{array}{c}
\begin{aligned}
\text{PD}_{{\text{T--W--ABORT--I}}|\beta} = \text{Pr}[\beta ^{\kappa-1}(1+{t_\text{GLRT--I}}) > {\eta _{\text{t}} } ;{\text{H}_1}]
= 1 -{{\cal P}_{1}{({\eta _{\text{t}}\beta ^{1-\kappa}}-1)}} ,
\end{aligned}
\end{array}
\end{equation}
respectively, where $\eta _{\text{a}}$, $\eta _{\text{w}}$, and $\eta _{\text{t}}$ are the detection thresholds for the ABORT--I, W--ABORT--I, and T--W--ABORT--I, respectively,
${{\cal P}_1}({\eta})$ is the cumulative distribution function (CDF) of   $t_\text{GLRT--I}$ in \eqref{GLRTHE} under hypothesis $\text{H}_1$ conditioned on ${\beta }$, given by
\begin{equation}
\label{CDF}
{\cal P}_1({\eta}) = \text{Pr} [t_\text{GLRT--I} \le {\eta }|{\beta }; \text{H}_1].
\end{equation}

Cautions must be taken when averaging the conditional PDs over $\beta$.  In \eqref{PDABORT}-\eqref{PDTWABORT}, to ensure that the CDF is meaningful, the following constraints are needed: $\beta\le\eta_{\text{a}}$, $\beta\le\eta_{\text{w}}$, and
\begin{equation}
\label{threshold}
\beta ^{1-\kappa}>\eta _{\text{t}}^{-1},
\end{equation}
respectively.
Consequently, together with the fact $0<\beta<1$, the expressions for the PDs of the ABORT--I and W--ABORT--I can be calculated as
\begin{equation}
\label{PDABORT1}
\begin{array}{c}
\begin{aligned}
\text{PD}{_{\text{ABORT--I}}} 
= \int_0^{\min(1,\eta_{\text{a}})} [1 -{{\cal P}_{1}{({\eta _{\text{a}}}-\beta)}}]{f_1}({\beta }) {\kern 1pt} \text{d}{\beta },
\end{aligned}
\end{array}
\end{equation}
and
\begin{equation}
\label{PDWABORT1}
\begin{array}{c}
\begin{aligned}
\text{PD}{_{\text{W--ABORT--I}}} 
= \int_0^{\min(1,\eta_{\text{w}})} \left[1 -{{\cal P}_{1}{\left( \frac{\eta _{\text{w}}}{\beta}-1   \right)}}\right]{f_1}({\beta }) {\kern 1pt} \text{d}{\beta },
\end{aligned}
\end{array}
\end{equation}
respectively. In \eqref{PDABORT1} and \eqref{PDWABORT1}, ${f_1}({\beta })$ is the probability density function (PDF) of ${\beta }$ defined in \eqref{Beta_HE} under hypothesis $\text{H}_1$.
The calculations of the PD of the T--W--ABORT--I are divided into the following four cases:
\begin{itemize}
	\item[i)] $0\le\kappa\le1$ and $\eta_{\text{t}}\le1$
	\begin{equation}
	\label{}
	\text{PD}{_{\text{T--W--ABORT--I}}}=1,
	\end{equation}
	
	\item[ii)] $0\le\kappa\le1$ and $\eta_{\text{t}}>1$
	\begin{equation}
	\label{}
	\text{PD}{_{\text{T--W--ABORT--I}}}=\int_{\eta_{\text{t}}^{-1/(1-\kappa)}}^1 [1 -{{\cal P}_{1}{({\eta _{\text{t}}\beta ^{1-\kappa}}-1)}}]{f_1}({\beta }) {\kern 1pt} \text{d}{\beta }，
	\end{equation}
	
	\item[iii)] $\kappa>1$ and $\eta_{\text{t}}\le1$
	\begin{equation}
	\label{}
	\text{PD}{_{\text{T--W--ABORT--I}}}=\int_0^{\eta_{\text{t}}^{-1/(1-\kappa)}} [1 -{{\cal P}_{1}{({\eta _{\text{t}}\beta ^{1-\kappa}}-1)}}]{f_1}({\beta }) {\kern 1pt} \text{d}{\beta },
	\end{equation}
	\item[iv)] $\kappa>1$ and $\eta_{\text{t}}>1$
	\begin{equation}
	\label{PDtwaborti4}
	\text{PD}{_{\text{T--W--ABORT--I}}}=\int_0^1 [1 -{{\cal P}_{1}{({\eta _{\text{t}}\beta ^{1-\kappa}}-1)}}]{f_1}({\beta }) {\kern 1pt} \text{d}{\beta }.	
	\end{equation}	
\end{itemize}	

In the presence of signal mismatch, $t_\text{GLRT--I} $ in \eqref{GLRTHE},
with a fixed $\beta $ under hypothesis $\text{H}_1$, is distributed as \cite{LiuLiu18AES}
\begin{equation}
\label{StDist_GLRT_HE_Mis}
t_\text{GLRT--I}|[\beta,\text{H}_1]\sim{\cal C}{{\cal F}_{p,L - N + q + 1}}({\rho_\text{eff}} {\beta }), 
\end{equation}
where
\begin{equation}
\label{rho_HE_eff_org}
\rho_\text{eff}={\bar \sbf_0}^H \Pbf_{\bar\Jbf}^\bot {\bar\Hbf}
(\bar\Hbf^H\Pbf_{\bar\Jbf}^\bot\bar\Hbf)^{-1}\bar\Hbf^H\Pbf_{\bar\Jbf}^\bot\bar \sbf_0
\end{equation}
is referred to as the effective signal-to-noise ratio (eSNR). In \eqref{rho_HE_eff_org}, $\bar \sbf_0=\Rbf^{-1/2}\sbf_0$,  $\bar \Jbf=\Rbf^{-1/2}\Jbf$,  $\bar \Hbf=\Rbf^{-1/2}\Hbf$, $\Pbf_{\bar \Jbf}^\bot=\Ibf_N-\Pbf_{\bar \Jbf}$, and $\Pbf_{\bar \Jbf}={\bar \Jbf}({\bar \Jbf}^H{\bar \Jbf})^{-1}{\bar \Jbf}^H$.
The statistical distribution of $t_\text{GLRT--I} $ in \eqref{GLRTHE} under hypothesis $\text{H}_0$ 
is  \cite{LiuLiu18AES}
\begin{equation}
\label{StDist_GLRT_HE_Misx}
t_\text{GLRT--I} \sim{\cal C}{{\cal F}_{p,L - N + q + 1}}, 
\end{equation}
Moreover,  in the presence of signal mismatch, $\beta $ in \eqref{Beta_HE} under hypotheses $\text{H}_1$ and $\text{H}_0$ is distributed as \cite{LiuLiu18AES}
\begin{equation}
\label{Beta_HE_H1}
\beta |\text{H}_1 \sim{\cal C}{{\cal B} _{L - N + p + q + 1,N - p - q}} (\delta^2 )
\end{equation}
and
\begin{equation}
\label{Beta_HE_H0}
\beta |\text{H}_0 \sim{\cal C}{{\cal B} _{L - N + p + q + 1,N - p - q}}, 
\end{equation}
respectively, where
\begin{equation}
\label{delta2_HE1}
\delta^2 =
{\bar \sbf_0^H \Pbf_{\bar\Jbf}^\bot  \Pbf_{\Pbf_{\bar\Jbf}^\bot\bar\Hbf}^\bot	\Pbf_{\bar\Jbf}^\bot   {\bar\sbf_0}}.
\end{equation}
with $\Pbf_{\Pbf_{\bar\Jbf}^\bot\bar\Hbf}^\bot=\Ibf_N-\Pbf_{\Pbf_{\bar\Jbf}^\bot\bar\Hbf}$ and
$\Pbf_{\Pbf_{\bar\Jbf}^\bot\bar\Hbf}=\Pbf_{\bar\Jbf}^\bot\bar\Hbf
(\bar\Hbf^H\Pbf_{\bar\Jbf}^\bot\bar\Hbf)^{-1}\bar\Hbf^H\Pbf_{\bar\Jbf}^\bot$.

According to (A2-29) in \cite{KellyForsythe89}, the CDF in \eqref{CDF} can be calculated as
\begin{equation}
\label{CDF111}
\begin{array}{l}
\begin{aligned}
{{\cal P}_{ 1}}({\eta}) =& \sum\limits_{k = 0}^{L - N + q} {\text{C}_{L-N+p+q}^{k + p}} \frac{{{\eta ^{k + p}}}}{{{{(1 + {\eta})}^{L-N+p+q}}}}{\text{IG}_{k + 1}}\left(\frac{{\rho_{\text{eff}} }\beta  } {1 + {\eta } }\right),
\end{aligned}
\end{array}
\end{equation}
where $\text{C}_n^m = \frac{n!}  {m!(n - m)!}$
is the binominal coefficient and
$\text{IG}{_{k + 1}}(a) = {e^{ - a}}\sum\nolimits_{m = 0}^k {\frac{{a^m}} {m!}}$
is the incomplete Gamma function.
Moreover, according to (A2-23) in \cite{KellyForsythe89}, the PDF of $\beta$ in \eqref{Beta_HE} under hypothesis $\text{H}_1$ is
\begin{equation}
\label{PDF_Beta_HE_Mis_1}
f_1(\beta )=f_{0}(\beta)\text{e}^{-\delta^2\beta}
\sum_{k=0}^{L-N+p+q+1}\text{C}_{L-N+p+q+1}^k\dfrac{(N-p-q-1)!}{(N-p-q+k-1)!} \delta^{2k}(1-\beta)^k,
\end{equation}
where 
\begin{equation}
\label{PDF_beta0}
f_{0}(\beta)= \frac{{{\beta ^{L-N+p+q}}{{(1 - \beta )}^{N-p-q-1}}}}{\text{B}({{L-N+p+q+1},N-p-q})}.
\end{equation}
is the PDF of $\beta$ under hypothesis $\text{H}_0$.
In \eqref{PDF_beta0}, 
${\text{B}({m,n})}=\frac{(m-1)!(n-1)!}{(m+n-1)!}$
is the Beta function. 
Taking \eqref{CDF111} and \eqref{PDF_Beta_HE_Mis_1} into \eqref{PDABORT1}--\eqref{PDtwaborti4}, we can obtain the final expression for the PDs of the ABORT--I, W--ABORT--I, and T--W--ABORT--I.

Setting $\rho_\text{eff}=0$ in \eqref{CDF111} results in the CDF of $t_\text{GLRT--I}$ under hypothesis $\text{H}_0$, i.e.,
\begin{equation}
\label{CDF00}
\begin{array}{l}
\begin{aligned}
{{\cal P}_{0}}({\eta}) =  {\text{C}_{L-N+p+q}^{ p}} \frac{\eta ^{ p}}{{(1 + {\eta})}^{L-N+p+q}} .
\end{aligned}
\end{array}
\end{equation}
The PFAs of the ABORT--I, W--ABORT--I, and T--W--ABORT--I can be obtained by replacing ${\cal P}_{1}(\cdot)$ and $f_1(\beta)$ by ${\cal P}_{0}(\cdot)$ and $f_0(\beta)$, respectively, in  \eqref{PDABORT1}--\eqref{PDtwaborti4}. 

Some remarks on the influence of signal mismatch on the detection performance of the detectors are given below. The eSNR in \eqref{rho_HE_eff_org} can be recast as \cite{LiuLiu18AES}
\begin{equation}
\label{rho_HE_eff_2}
\rho_\text{eff}=\rho_\text{SNR}\sin^2\psi\cos^2\vartheta,
\end{equation}
where
\begin{equation}
\label{}
\rho_\text{SNR}={\bar \sbf_0}^H{\bar \sbf_0}
\end{equation}
is the conventional SNR for multichannel signal detection in the absence of interference,
\begin{equation}
\label{}
\sin^2\psi=\frac{{\bar \sbf_0}^H \Pbf_{\bar\Jbf}^\bot{\bar \sbf_0} }{{\bar \sbf_0}^H {\bar \sbf_0} },
\end{equation}
and
\begin{equation}
\label{cos2vartheta}
\cos^2\vartheta
=\frac{\bar \sbf_0^H{\Pbf_{\Pbf_{\bar\Jbf}^\bot\bar \Hbf}}\bar \sbf_0}{{\bar \sbf_0}^H \Pbf_{\bar\Jbf}^\bot{\bar \sbf_0} }.
\end{equation}
The quantity $\cos^2\vartheta$ in \eqref{cos2vartheta} serves as the metric of signal mismatch in the presence of interference. If signal mismatch does not occur, there exists a $p\times1$ vector $\thetabf_0$ such that $\sbf_0=\Hbf\thetabf_0$. Using this result, we can verify that $\cos^2\vartheta=1$.

For comparison purposes, a well-known metric of signal mismatch in the absence of interference is listed below \cite{LiuLiu16SP}
\begin{equation}
\label{cos2psi}
\cos^2\phi=\frac{\bar \sbf_0^H\Pbf_{\bar\Hbf}\bar \sbf_0} {{\bar \sbf_0}^H{\bar \sbf_0} }.
\end{equation}
Some preliminary analysis is summarized in the following proposition.

\textbf{\textit{Proposition 1.}}  i). $\cos^2\phi=1$ results in $\cos^2\vartheta=1$, but not vice versa. ii) $\cos^2\phi=0$  does not necessarily lead to $\cos^2\vartheta=0$, and vice versa.

\noindent
\textbf{{Proof.}}  See the appendix A. $\hfill{} \blacksquare$

Before proceeding, we would like to point out that the three proposed detectors can successfully suppress the interference, since the power of the interference does not impact the PDs and PFAs. The interference 
affects the detection performance through $\sin^2\psi$ and the DOFs of the statistical distributions. More analysis of the influence of interference on the detection performance can be found in \cite{LiuLiu18AES}.


\section{Numerical examples}
In this section, we evaluate the detection performance of the proposed ABORT--I, W--ABORT--I,  and T--W--ABORT--I for the case of no signal mismatch and the case of signal mismatch. Both theoretical and Monte Carlo simulation results are provided. The noise is modelled as exponentially correlated random vector with one-lag correlation coefficient. Hence, the $(i,j)$th element of $\Rbf$ is $\Rbf(i,j)=\epsilon^{|i-j|}$, $i,j=1,2,\cdots,N$, and  $\epsilon$ is chosen to be $0.9$.  The interference-to-noise ratio (INR) is defined as
\begin{equation}
\label{INR}
\text{INR}=\phibf^H\Jbf^H\Rbf^{-1}\Jbf\phibf.
\end{equation}
To 
reduce the running time of Monte Carlo simulations, the PFA is chosen as $\text{PFA}=10^{-3}$.  $ 10^{5}$ Monte Carlo simulations are used to generate a detection threshold, while $ 10^{4}$ Monte Carlo simulations are carried out to generate a PD. Moreover, the following parameters are adopted throughout this section: $N=12$, $L=2N$, $p=1$, $q=2$, and $\text{INR}=10~\text{dB}$.

Fig. 1 
shows the PDs of the proposed detectors under different SNRs,  compared with the existing GLRT--I and 2S--GLRT--I. 
In the legend, ``TH'' indicates theoretical results, while ``MC'' stands for Monte Carlo simulation results.
It is seen that the theoretical results match the Monte Carlo simulation results pretty well. For the chosen parameters, the ABORT--I, T--W--ABORT--I with $\kappa=0.8$, GLRT--I, and 2S--GLRT--I roughly have the same PDs. The W--ABORT--I and T--W--ABORT--I with $\kappa=2.5$ suffer from certain performance loss for matched signals, compared with the other detectors. However, the W--ABORT--I and T--W--ABORT--I with $\kappa=2.5$ exhibit satisfied detection performance in terms of rejecting mismatch signals, as shown in Fig. 3 below.

Fig. 2 plots the PDs of the detectors under different $\sin^2\psi$. The tunable parameter for the T--W--ABORT-I is $\kappa=0.8$. The PD curve of the W--ABORT--I is not given, since it suffers from certain detection performance loss, compared with the other detectors.
The results show that all the PDs of the detectors increase when $\sin^2\psi$ increases. This is because the increase of $\sin^2\psi$ results in the increase of the eSNR, defined in \eqref{rho_HE_eff_org}, which leads to the improvement in the PD. 

Fig. 3 
displays the PDs of the T--W--ABORT--I with different tunable parameters $\kappa$. It is shown that when there is no signal mismatch, i.e., $\cos^2\vartheta=1.0$, the PD of the T--W--ABORT--I first increases and then decreases as the tunable parameter $\kappa$ increases.  In contrast, when signal mismatch occurs, i.e., $\cos^2\vartheta=0.3$, the PD of the T--W--ABORT--I decreases directly as the tunable parameter $\kappa$ increases. This is due to the fact that the selectivity of the T--W--ABORT--I increases as the increase of the tunable parameter.
Specifically, in the range of $0.6\le\kappa\le1.0$,  the T--W--ABORT--I can provide roughly the same PD as the GLRT--I (a special case of the T--W--ABORT--I with $\le\kappa=1.0$) in the case of no signal mismatch.

Fig. 4 
depicts the contours of the PDs of the detectors under different degrees of signal mismatch and different SNRs.  This type of figure is usually called mesa plot. The solid lines denote theoretical results, which are consistent with the Monte Carlo results indicated by the dotted lines.
It is shown that the ABORT--I and W--ABORT-I have better detection performance than the GLRT--I and 2S--GLRT--I in terms of mismatched signal rejection. 
Taking the ABORT--I for example. When $\cos^2\vartheta<0.5$, it cannot provide a PD greater than 0.5, no matter how high the SNR is.
In other words, the ABORT--I and W--ABORT-I do not take a largely mismatched signal as a desired target.
Moreover, the T--W--ABORT--I is very flexible in controlling the detection performance for mismatched signals. With a large tunable parameter, i.e., $\kappa=2.5$, the T--W--ABORT--I possesses the best selectivity property. On the other hand, the T--W--ABORT--I, with a small tunable parameter, is very robust to signal mismatch. For the chosen parameters, the T--W--ABORT--I with $\kappa=0.8$ roughly has the same robustness as the 2S--GLRT--I. In fact, the T--W--ABORT--I with a smaller tunable parameter can become much more robust than the 2S--GLRT--I.

Gathering the results in Figs. 1, 2, and 4, we can conclude that: 1) The ABORT--I has slightly better selectivity property than the GLRT--I. However, the former suffers from slightly performance loss compared with the later in the case of no signal mismatch.
2) The T--W--ABORT--I, with a proper tunable parameter less than unity, can provide better robustness than the GLRT--I and 2S--GLRT--I. The T--W--ABORT--I, with the same tunable parameter, suffers from a slightly performance loss for matched signals, compared with the GLRT--I.
3) The W--ABORT--I and T--W--ABORT--I with a proper tunable parameter greater than two are much more selective than the other detectors. However, these two detectors suffer from non-negligible loss in the case of no signal mismatch.

\section{Conclusions}
In this paper, we considered the problem of detecting a multichannel signal in the presence of interference when signal mismatch happens. Two selective detectors, namely, the ABORT--I and W--ABORT--I, and a tunable detector, namely, T--W--ABORT--I were proposed, and the corresponding analytical expressions for the PDs and PFAs were given. Numerical examples show that the ABORT--I and W--ABORT--I exhibit better detection performance in terms of rejecting mismatched signals, and the T--W--ABORT--I has the flexibility in governing the detection performance for mismatched signals. The T--W--ABORT--I, with a large tunable parameter, is very selective, while it becomes robust 
with a moderately small tunable parameter. In addition, in the case of no signal mismatch, the ABORT--I and T--W--ABORT--I with a suitable tunable parameter, say, $0.6\le\kappa\leq1.0$, can provide nearly the same detection performance as the GLRT--I.

\appendix
\section{Proof of Proposition 1}
i) If $\cos^2\phi=1$, then there exists a $p\times1$ vector $\thetabf_0$ such that $\bar\sbf_0=\bar\Hbf\thetabf_0$. Taking this result into \eqref{cos2vartheta} yields that $\cos^2\vartheta=1$.

On the other hand, if $\cos^2\vartheta=1$, then we have 
\begin{equation}
\label{}
{\Pbf_{\Pbf_{\bar\Jbf}^\bot\bar \Hbf}}\bar \sbf_0=\Pbf_{\bar\Jbf}^\bot\bar \sbf_0,
\end{equation}
which can be recast as
\begin{equation}
\label{Ppjvhs}
{\Pbf_{\Pbf_{\bar\Jbf}^\bot\bar \Hbf}}\Pbf_{\bar\Jbf}^\bot\bar \sbf_0=\Pbf_{\bar\Jbf}^\bot\bar \sbf_0.
\end{equation}
It follows that $\Pbf_{\bar\Jbf}^\bot\bar \sbf_0$ lies in the subspace 
$<{\Pbf_{\bar\Jbf}^\bot\bar \Hbf}>$. Hence, there exists a $p\times1$ vector $\thetabf_1$ such that
\begin{equation}
\label{Ppjvhsxx}
{{\Pbf_{\bar\Jbf}^\bot\bar \Hbf}}\thetabf_1=\Pbf_{\bar\Jbf}^\bot\bar \sbf_0.
\end{equation}
Using the matrix $\bar\Jbf$, we can obtain an $N\times(N-q)$ semi-unitary matrix $\bar\Jbf_{_\bot}$ such that
\begin{equation}
\label{Pjbf}
\Pbf_{\bar\Jbf}^\bot=\bar\Jbf_{_\bot}\bar\Jbf_{_\bot}^H,
\end{equation}
$\bar\Jbf_{_\bot}^H\bar\Jbf_{_\bot}=\Ibf_{N-q}$, and
\begin{equation}
\label{JvJ}
\bar\Jbf_{_\bot}^H\bar\Jbf=\mathbf{0}_{(N-q)\times q}.
\end{equation}
Then \eqref{Ppjvhsxx} can be rewritten as
\begin{equation}
\label{Ppjvhsx1x}
\bar\Jbf_{_\bot}\bar\Jbf_{_\bot}^H\bar \Hbf\thetabf_1=\bar\Jbf_{_\bot}\bar\Jbf_{_\bot}^H\bar \sbf_0.
\end{equation}
According to \eqref{JvJ}, \eqref{Ppjvhsx1x} can be rewritten as
\begin{equation}
\label{Ppjvhsx1xx}
\bar\Jbf_{_\bot}\bar\Jbf_{_\bot}^H(\bar \Hbf\thetabf_1+\bar\Jbf\phibf_1)=\bar\Jbf_{_\bot}\bar\Jbf_{_\bot}^H\bar \sbf_0,
\end{equation}
where $\phibf_1$ is an arbitrary $q\times1$ vector.
It follows from \eqref{Ppjvhsx1xx} that if the whitened signal component $\bar\sbf_0$ can be expressed as
\begin{equation}
\label{s0s0}
\sbf_0=\bar \Hbf\thetabf_1+\bar\Jbf\phibf_1,
\end{equation}
then $\cos^2\vartheta=1$. It is known from \eqref{s0s0} that $\bar\sbf_0$ may not completely lie in $<\bar\Hbf>$ when $\cos^2\vartheta=1$. In this case, we have $\cos^2\phi<1$. 

ii) If $\cos^2\phi=0$, then $\bar\Hbf^H\bar\sbf_0=\textbf{0}_{p\times1}$, or equivalently,
\begin{equation}
\label{Hbfsbf0}
\bar\Hbf_{_{//}}^H\bar\sbf_0=\textbf{0}_{p\times1},
\end{equation}
where
$\bar\Hbf_{_{//}}=\bar\Hbf(\bar\Hbf^H\bar\Hbf)^{-\frac{1}{2}}$.
Using $\bar\Hbf_{_{//}}$ we can construct an $N\times N$ unitary matrix $\Ubf=[\bar\Hbf_{_{//}},\bar\Hbf_{_{\bot}}]$,
which can be taken as a basic matrix of the   $N\times N$ complex space $\mathbb{C}^{N\times N}$. Hence, there exists an $N\times 1$ vector $\underline\thetabf$ such that
\begin{equation}
\label{Ubftheta}
\bar\sbf_0= \Ubf\underline\thetabf.
\end{equation}
We can partition $\underline\thetabf$ as $\underline\thetabf=[\thetabf_{_{//}}^T,\thetabf_{_{\bot}}^T]^T$, where the dimensions of $\thetabf_{_{//}}$ and $\thetabf_{_{\bot}}$ are $p\times1$ and $(N-p)\times1$, respectively. 
According to the definitions of $\Ubf$ and $\underline{\thetabf}$, we have
\begin{equation}
\label{Ubftheta2}
\bar\sbf_0=\bar\Hbf_{_{//}}\thetabf_{_{//}}+\bar\Hbf_{_{\bot}}\thetabf_{_{\bot}}.
\end{equation}
Pre-multiplying \eqref{Ubftheta2} with $\bar\Hbf_{_{//}}^H$ yields that
\begin{equation}
\label{theta2}
\thetabf_{_{//}}=\textbf{0}_{p\times1}.
\end{equation}
Substituting \eqref{theta2} into \eqref{Ubftheta2} leads to
\begin{equation}
\label{Ubftheta3}
\bar\sbf_0=\bar\Hbf_{_{\bot}}\thetabf_{_{\bot}}.
\end{equation}
Substituting \eqref{Pjbf} and \eqref{Ubftheta3} into \eqref{cos2vartheta} leads to
\begin{equation}
\label{cos2vartheta2}
\cos^2\vartheta=\frac{\thetabf_{_{\bot}}^H\bar\Hbf_{_{\bot}}^H \bar\Jbf_{_{\bot}}\bar\Jbf_{_{\bot}}^H\bar\Hbf_{_{//}} (\bar\Hbf_{_{//}} ^H\bar\Jbf_{_{\bot}}\bar\Jbf_{_{\bot}}^H\bar\Hbf_{_{//}} )^{-1}\bar\Hbf_{_{//}}^H\bar\Jbf_{_{\bot}}\bar\Jbf_{_{\bot}}^H	 \bar\Hbf_{_{\bot}}\thetabf_{_{\bot}}}
{\thetabf_{_{\bot}}^H\bar\Hbf_{_{\bot}}^H \bar\Jbf_{_{\bot}}\bar\Jbf_{_{\bot}}^H \bar\Hbf_{_{\bot}}\thetabf_{_{\bot}}},
\end{equation}
which is generally not equal to zero. For example, a specific form of $\bar\Jbf_{_{\bot}}$, for a given nominal signal matrix $\Hbf$, is
$\bar\Jbf_{_{\bot}}=[\bar\Hbf_{_{//}},\bar\Hbf_{_{\bot},1}]$, where $\bar\Hbf_{_{\bot},1}$ is the first $N-p-q$ columns of $\bar\Hbf_{_{\bot}}$.

If $\cos^2\vartheta=0$, then
\begin{equation}
\label{HPjvs0}
\bar\Hbf^H\Pbf_{\bar\Jbf}^\bot\bar\sbf_0=\textbf{0}_{p\times1}.
\end{equation}
In a manner similar to the derivations of \eqref{Ubftheta}-\eqref{Ubftheta3}, $\bar\sbf_0$ can be expressed as
\begin{equation}
\label{Vbftheta3}
\bar\sbf_0=\bar\Vbf\thetabf_2,
\end{equation}
where $\bar\Vbf$ is an $N\times(N-p)$ matrix such that the augmented matrix $[\Pbf_{\bar\Jbf}^\bot\bar\Hbf(\bar\Hbf^H\Pbf_{\bar\Jbf}^\bot\bar\Hbf)^{-\frac{1}{2}}, \bar\Vbf]$ is an $N\times N$ unitary matrix, and $\thetabf_2$ is an $(N-p)\times1$ vector.
Substituting \eqref{Vbftheta3} into \eqref{cos2psi} results in
\begin{equation}
\label{cos2psi2}
\cos^2\phi=\frac{\thetabf_2^H\bar \Vbf^H\Pbf_{\bar\Hbf}\bar \Vbf\thetabf_2} {\thetabf_2^H\bar \Vbf^H\bar \Vbf\thetabf_2 }.
\end{equation}
Another form of \eqref{cos2psi2} is given below. It is known from \eqref{HPjvs0} that
\begin{equation}
\label{HPjvs0x}
\bar\Hbf^H\bar\sbf_0=\bar\Hbf^H\Pbf_{\bar\Jbf}\bar\sbf_0.
\end{equation}
Substituting \eqref{HPjvs0x} into \eqref{cos2psi}, after some algebra, leads to
\begin{equation}
\label{cos2psi2x}
\cos^2\phi=\frac{\bar\sbf_0^H\Pbf_{\bar\Jbf}\Pbf_{\bar\Hbf}\Pbf_{\bar\Jbf}\bar\sbf_0} {\bar\sbf_0^H\bar\sbf_0 }.
\end{equation}
From \eqref{cos2psi2} and \eqref{cos2psi2x}, we know that $\cos^2\phi$ is not generally equal to zero.

\section{The method to generate $\Jbf$, $\jbf$, $\sin^2\psi$, and $\cos^2\vartheta$ with specific values for Monte Carlo simulations}

There are three main steps to generate  $\Jbf$, $\sin^2\psi$, and $\cos^2\vartheta$ with specific values:
1) Generate an arbitrary $N\times q$ matrix $\Jbf$. 
2) Generate the actual signal steering vector $\sbf_0$ satisfying a specific value of $\sin^2\psi$. Precisely, $\sbf_0$ is generated by selecting a properly scalar $0\leq r\leq 1$ such that $\sbf_0=\Rbf^{\frac{1}{2}}\bar\sbf_0$ and $\bar\sbf_0=r \bar \jbf_0+(1-r)\bar\jbf_1$ satisfying a specific $\sin^2\psi$, with $\bar \jbf_0$ being an arbitrary column of $\bar\Jbf$ and $\bar\jbf_1$ being the last column of $\Abf$. $\Abf$ is the matrix containing the left singular-vectors of ${\bar\Jbf}$.
3) Generate the nominal signal matrix $\Hbf$ satisfying specific value of $\cos^2\vartheta$.
Precisely, $\Hbf$ 
can be generated by choosing an appropriate scalar $0\le \alpha\le 1$ such that $\Hbf=\Rbf^{\frac{1}{2}}\bar\Hbf$ and  $\bar\Hbf=\alpha\bar\Hbf_0+(1-\alpha)\bar\Hbf_1$ satisfies a specific $\cos^2\vartheta$, with $\bar\Hbf_0=[\bar\sbf_0,\bar\Hbf_r]$ and $\bar\Hbf_1=\Wbf_1$. $\bar\Hbf_r$ is an arbitrary $N\times(p-1)$ matrix, and $\Wbf_1$ is the last $p$ columns of $\Wbf$, with $\Wbf$ containing the left singular-vectors of $\Pbf_{\bar\Jbf}^\bot\bar\sbf_0$.

Moreover, for a given INR defined in (43), 
we can generate the interference $\jbf$ as $\jbf=c\Jbf\phibf_n$, where $\phibf_n$ is an arbitrary $q\times1$ column vector and $c=\sqrt{\phibf_n^H\Jbf^H\Rbf^{-1}\Jbf\phibf_n}$.

\section*{Acknowledgements}
This work was supported in part by National Natural Science Foundation of China under Contracts 61501505, 61501351, and 61871469, in part by the Natural Science Foundation of Hubei Province under Contract 2017CFB589, in part by the National Natural Science Foundation of China and Civil Aviation Administration of China under Grant U1733116, and in part by the Youth Innovation Promotion Association CAS under Grant CX2100060053.

\section*{References}
\bibliographystyle{IEEEtran}
\bibliography{Detection}

\clearpage

\begin{figure}[htbp] 
	\setlength{\abovecaptionskip}{2pt}
	\setlength{\belowcaptionskip}{2pt}
	\centering
	\includegraphics[width=1\textwidth]{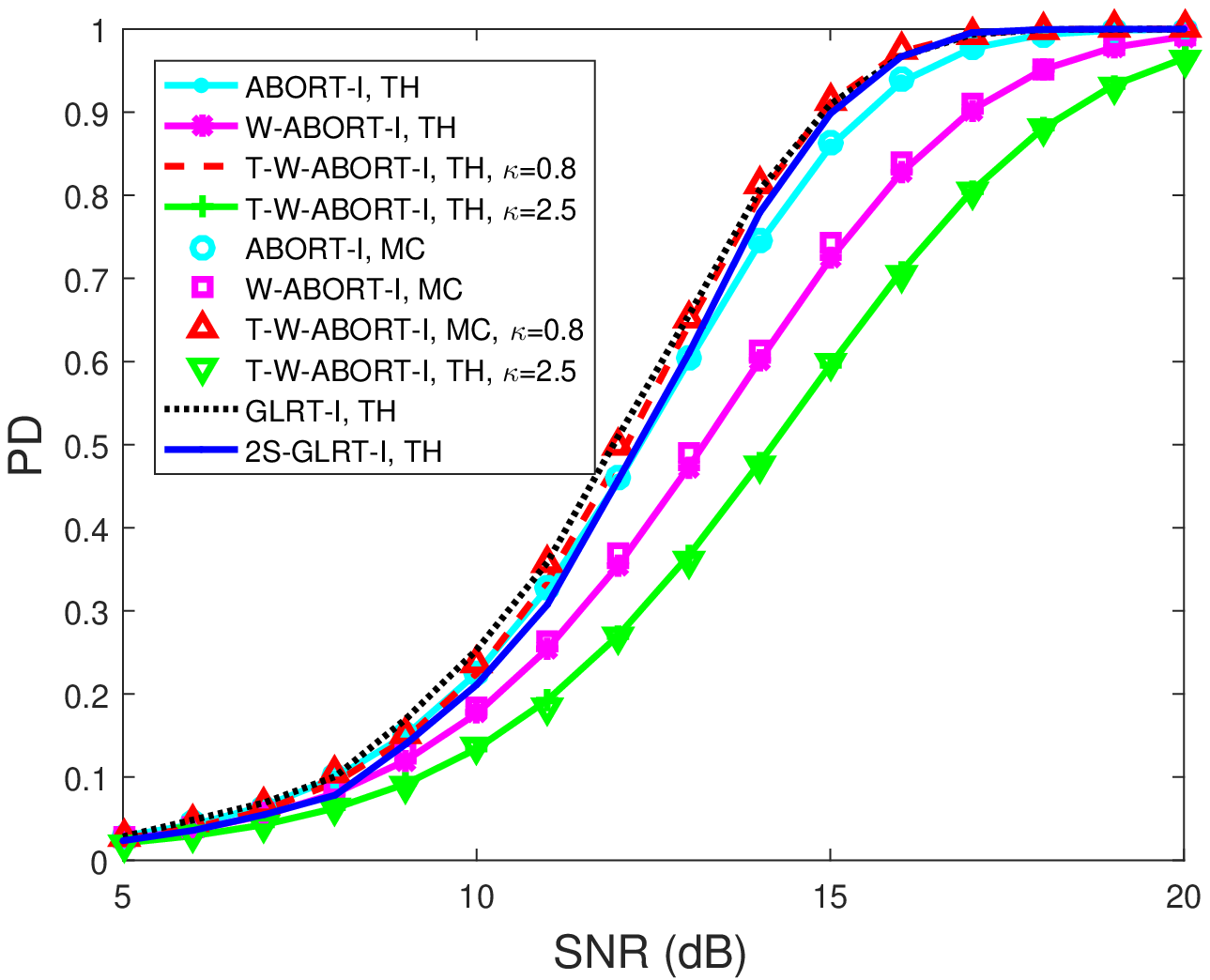}
\\
{Fig.1.~~PD versus SNR in the absence of signal mismatch.   $\cos^2\vartheta=1$ and $\text{sin}{^2}\psi =0.8$.}

	\label{PDvsSNRN12L1d5Np1q2}
\end{figure}

\begin{figure}[htbp] 
	\setlength{\abovecaptionskip}{2pt}
	\setlength{\belowcaptionskip}{2pt}
	\centering
	\includegraphics[width=1\textwidth]{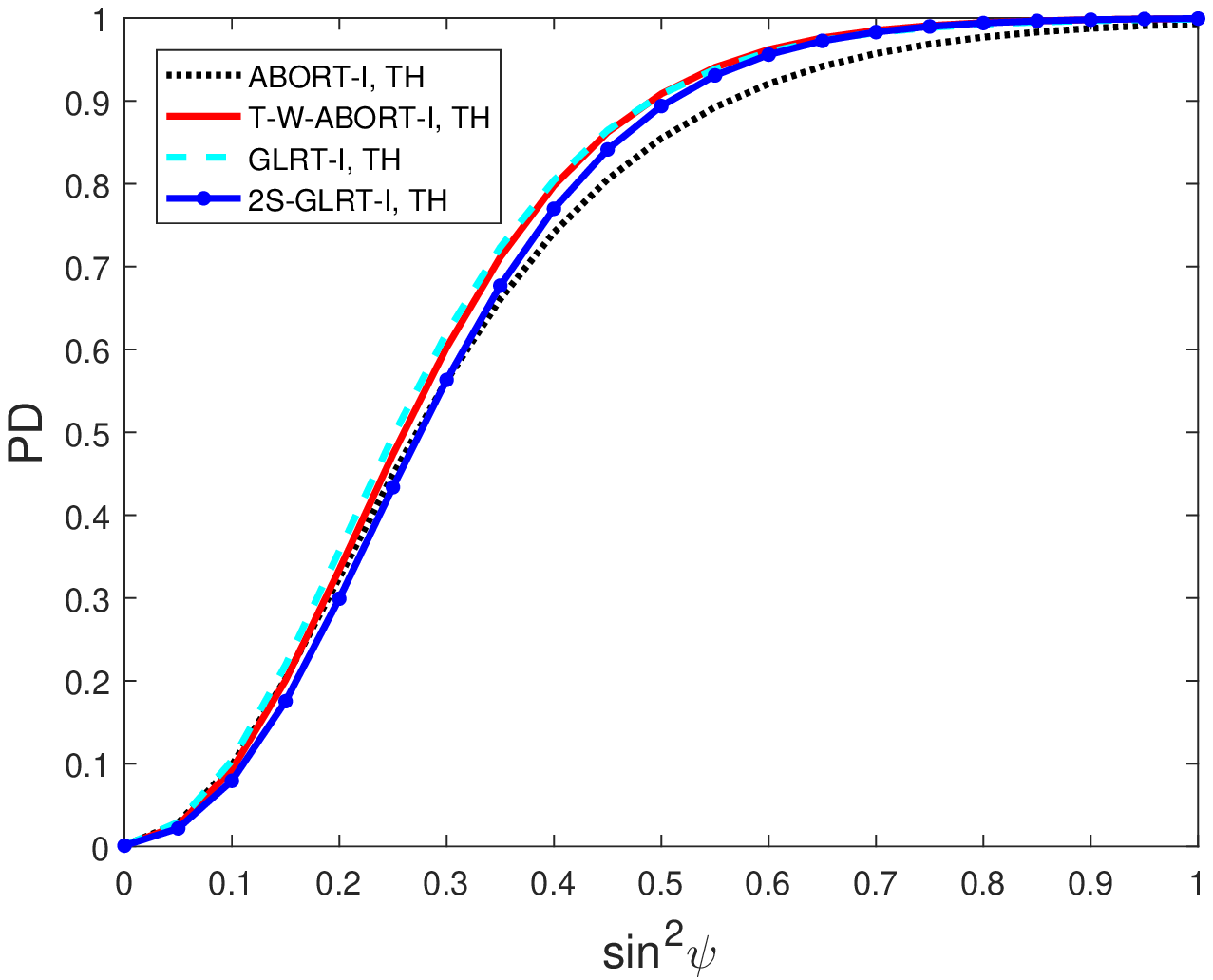}
\\ {Fig. 2. ~~ PD versus $\text{sin}{^2}\psi$ in the absence of signal mismatch. $\cos^2\vartheta=1$ and $\text{ SNR}=17~\text{ dB}$.}
	\label{PDvsSNRN12L1d5Np1q2}
\end{figure}

\begin{figure}[htp]
	\setlength{\abovecaptionskip}{2pt}
	\setlength{\belowcaptionskip}{2pt}
	\centering
	\includegraphics[width=1\textwidth]{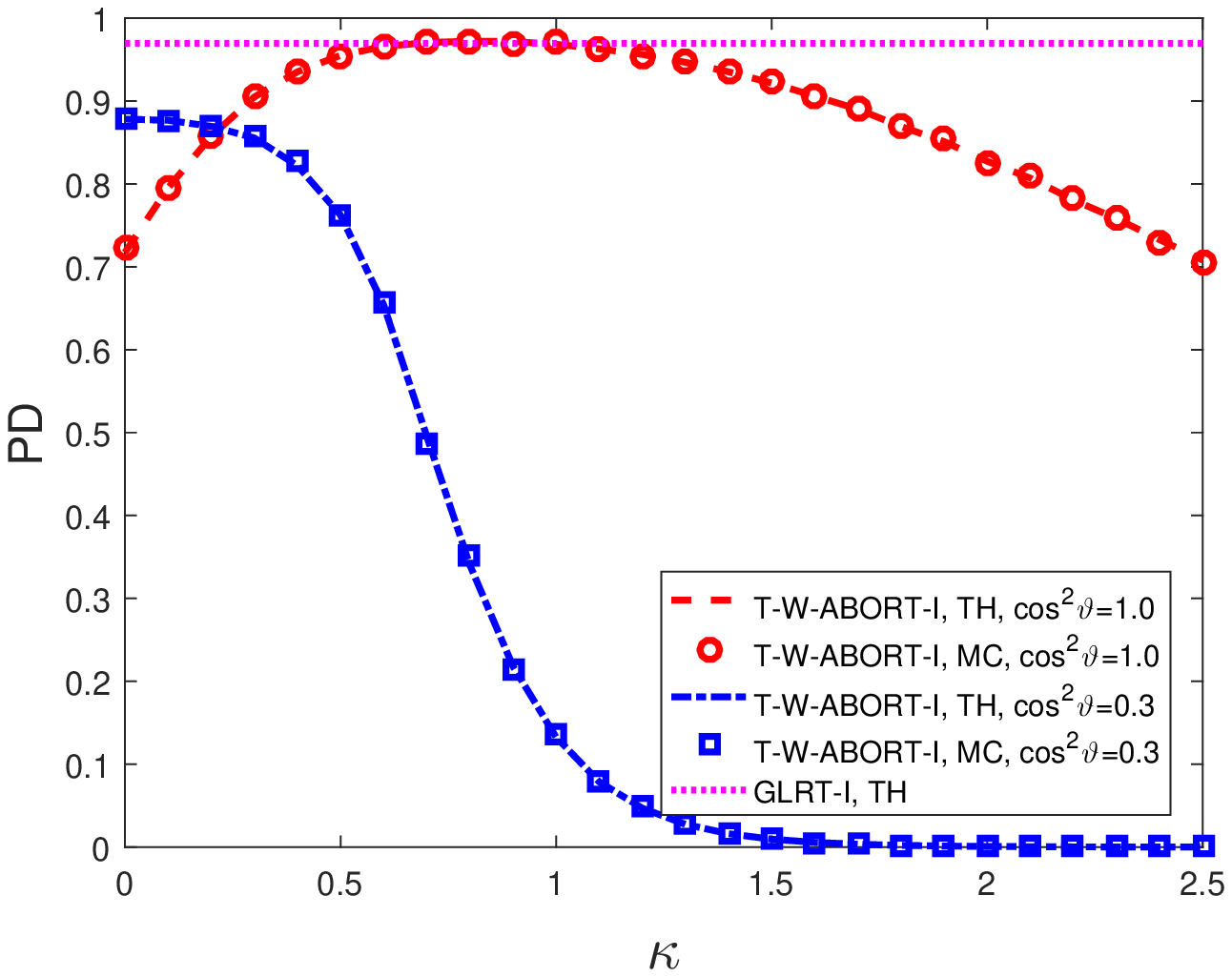} 
\\{Fig. 3. ~~PD of the T--W--ABORT--I versus $\kappa$. $\text{sin}{^2}\psi =0.8$ and $\text{ SNR}=17~\text{ dB}$.}
	\label{PDvsKappaN12L2Np1q2}
\end{figure}

\begin{figure}[htbp]
	\setlength{\abovecaptionskip}{2pt}
	\setlength{\belowcaptionskip}{2pt}
	\centering
	\includegraphics[width=0.7\textwidth]{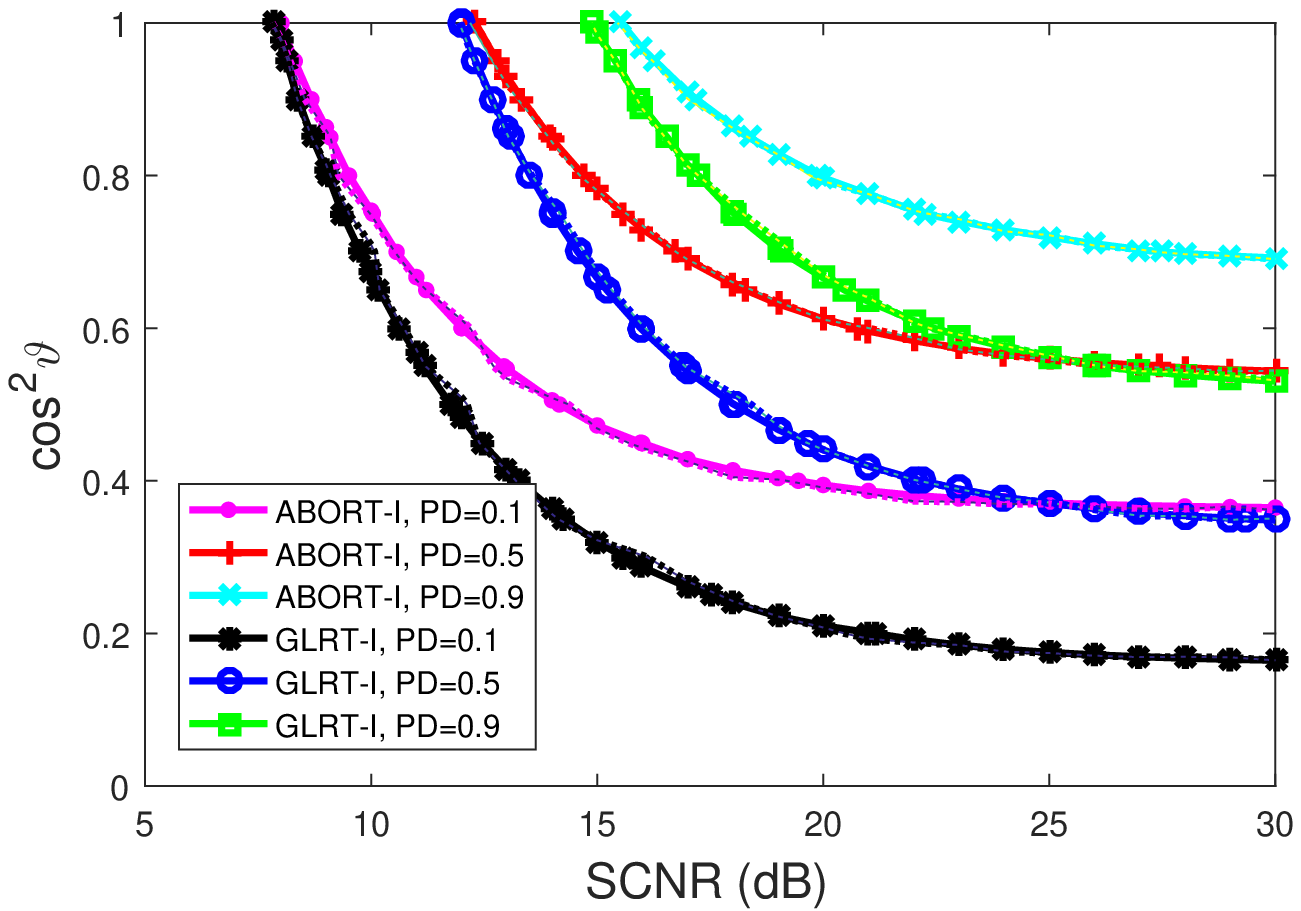} \\
	\includegraphics[width=0.7\textwidth]{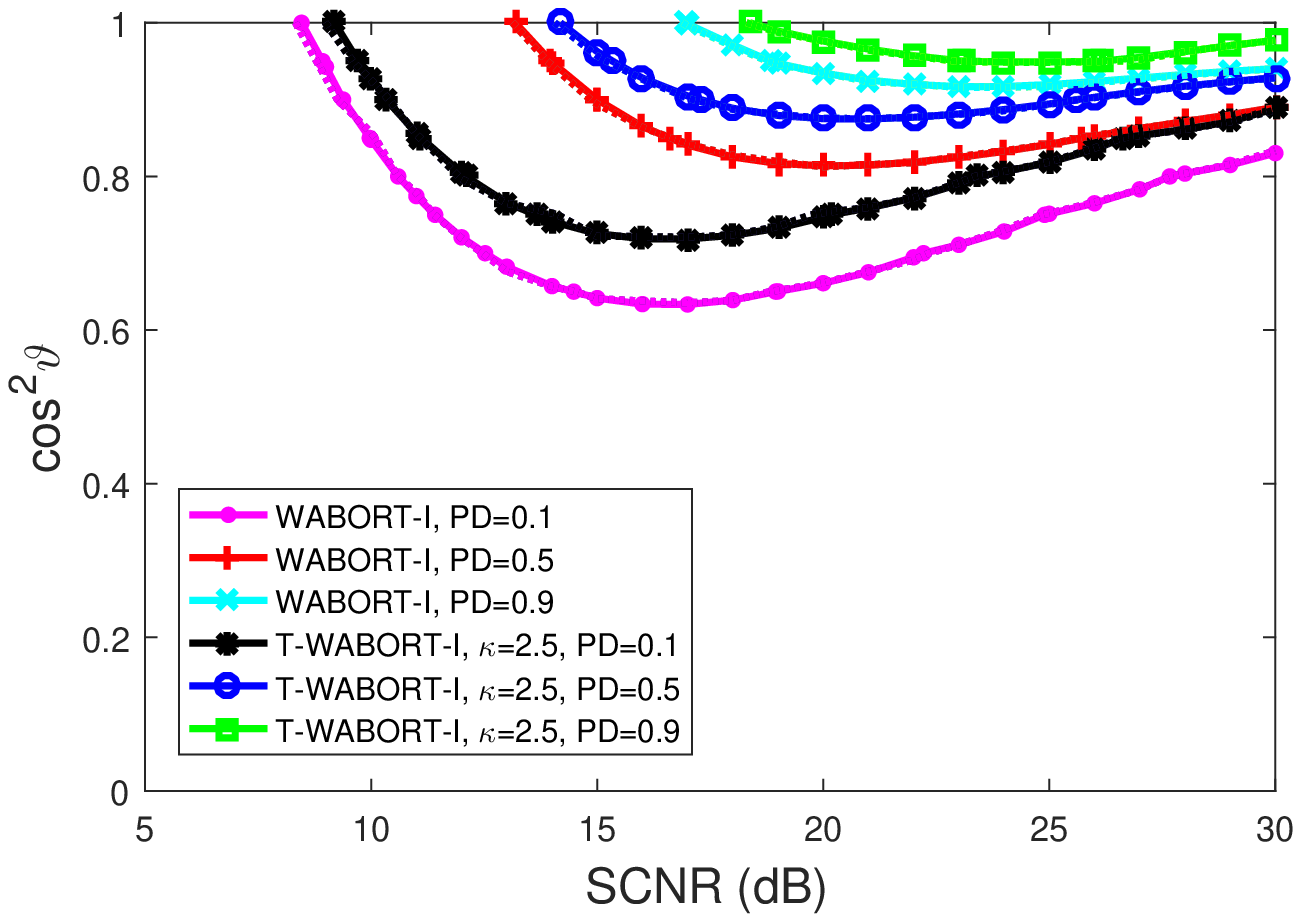} \\
	\includegraphics[width=0.7\textwidth]{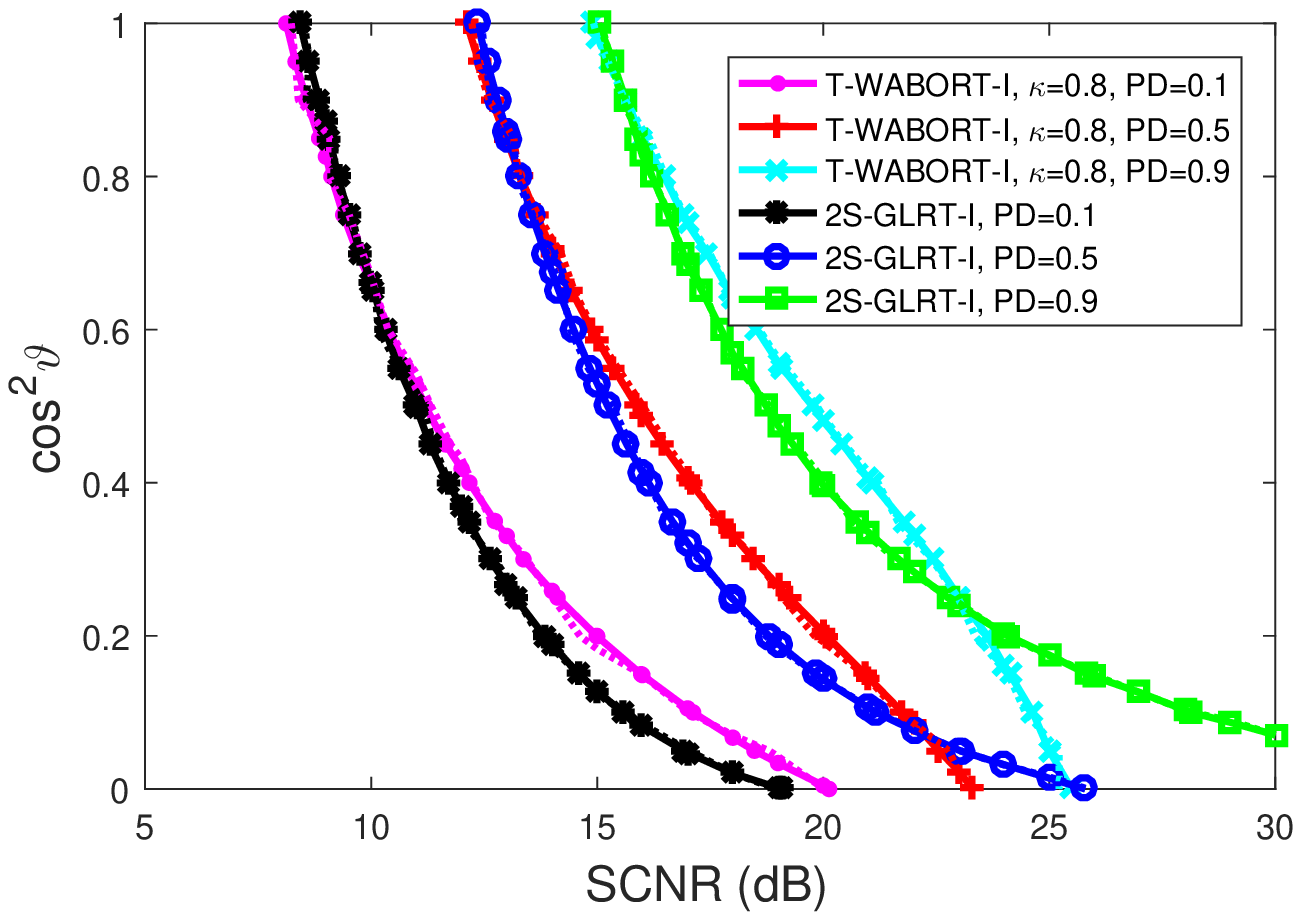} \\
{Fig. 4. ~~Contours of  PDs vs SNR and  $\text{cos}{^2}\vartheta$.  \ \ $\text{sin}{^2}\psi=0.8$. The solid lines with symbols denote theoretical results, while the dotted lines stand for the Monte Carlo results.}
	\label{PDvscos2varthetaN12L2Np1q2}
\end{figure}

\end{document}